\documentstyle[bezier,emlines2,11pt]{article}
\title{\bf
Conformal invariant equations for nucleus-nucleus scattering in
perturbative QCD with $N_c\to\infty$}
\author{M.Braun \\ Department of High Energy physics,
 University of S.Petersburg,\\
198504 S.Petersburg, Russia}
\def\beq{\begin{equation}}
\def\eeq{\end{equation}}
\def\noi{\noindent}

\begin{document}

\maketitle

\noi{\bf Abstract.}
In the perturbative QCD with $N_c\rightarrow\infty$
the amplitude for the collision of two  heavy nuclei is
expressed via  dipole densities in the nuclei.
Coupled equations for these densities
are derived in the configuration space. The equations are conformal
invariant in absence of external sources. Passing to conformal basis and
its possible truncation  are discussed.

\section{Introduction}
In the perturbative QCD with $N_c\to\infty$ , high-energy  scattering
of a pointlike projectile on a large nucleus is described by a sum of
fan diagrams constructed from BFKL pomerons and verteces
for their splitting in two. Summation of these diagrams leads to the
well-known BK equation ~\cite {bal,kov,bra1}, now well studied both
theoretically and numerically. Generalization of these results to
nucleus-nucleus scattering requires a symmetric treatment of projectile
and target and obviously involves not only the vertex for the splitting
of a pomeron in two but also fusing of two pomerons in one. Such a
program was realized in our papers ~\cite{bra2}. There we limited
ourselves to the case when  momenta transferred to both nuclei are
neglegible as compared to  gluon momenta inside pomerons,
which physically  corresponds to the limit of very heavy nuclei and a finite
pomeron slope. As a result all pomerons were propagating in the forward
direction. This simplified the final equations considerably but their basic
conformal invariance property remained hidden. Also written in the
momentum space the equations are difficult to compare with  the results
following from  the dipole picture and so-called JIMWLK approach, in which
the problem of symmetric treatment of projectile and target (and also of
inclusion of pomeron loops) is lately being studied very actively
(see e.g. ~\cite{jimwlk} and references therein). For these reasons in this
paper we rederive equations
describing nucleus-nucleus scattering for a general case, when the pomerons
are allowed to change their momenta in their interaction with the nuclei and
between themselves. Our final equations are in the transverse
coordinate space, so that comparison with the dipole approach will be
facilitated.

It is important to stress the approximations used in the derivation.
We rely on the perturbative QCD  in the limit of large number
of colours, $N_c\to\infty$, and assume both nuclei to be large,
with their atomic numbers $A,B>>1$. Thich allows to take into account only
tree diagrams constructed of pomerons and their interaction verteces
(see Fig. 1 for the simplest examples).
\begin{figure}
\unitlength=1mm
\special{em:linewidth 0.4pt}
\linethickness{0.4pt}
\begin{picture}(130.27,138.67)
\put(16.00,120.33){\circle*{5.20}}
\put(41.00,129.33){\circle*{5.20}}
\put(33.67,110.00){\circle*{5.20}}
\put(47.00,110.00){\circle*{5.20}}
\put(63.33,129.33){\circle*{5.20}}
\put(80.67,129.33){\circle*{5.20}}
\put(104.00,129.33){\circle*{5.20}}
\put(127.67,129.33){\circle*{5.20}}
\put(104.00,110.00){\circle*{5.20}}
\put(127.67,110.33){\circle*{5.20}}
\put(114.67,120.00){\circle*{5.20}}
\put(15.67,138.33){\line(0,-1){15.00}}
\put(17.67,138.00){\line(0,-1){15.00}}
\put(15.67,117.67){\line(0,-1){16.67}}
\put(17.67,117.67){\line(0,-1){16.67}}
\put(40.00,138.00){\line(0,-1){6.33}}
\put(42.33,138.00){\line(0,-1){6.33}}
\put(40.00,126.67){\line(0,-1){6.00}}
\put(42.33,126.67){\line(0,-1){5.67}}
\put(40.00,120.67){\line(-4,-5){6.00}}
\put(42.33,120.67){\line(3,-4){6.00}}
\put(41.00,118.33){\line(-5,-6){5.00}}
\put(41.33,118.00){\line(5,-6){4.67}}
\put(31.67,107.67){\line(0,-1){6.33}}
\put(34.67,107.33){\line(0,-1){5.67}}
\put(45.67,107.67){\line(0,-1){6.33}}
\put(49.00,107.67){\line(0,-1){6.33}}
\put(61.67,138.33){\line(0,-1){6.33}}
\put(64.33,138.33){\line(0,-1){6.33}}
\put(71.00,119.67){\line(0,-1){7.00}}
\put(73.67,120.00){\line(0,-1){7.67}}
\put(62.00,126.67){\line(5,-4){9.00}}
\put(81.33,126.67){\line(-6,-5){7.67}}
\put(65.33,127.00){\line(5,-4){7.00}}
\put(72.00,121.67){\line(6,5){6.67}}
\put(79.33,138.67){\line(0,-1){7.33}}
\put(82.33,138.67){\line(0,-1){7.00}}
\put(72.33,110.00){\circle*{5.20}}
\put(71.00,107.67){\line(0,-1){6.67}}
\put(73.67,107.67){\line(0,-1){6.67}}
\put(102.00,138.67){\line(0,-1){7.33}}
\put(102.00,126.67){\line(0,-1){14.00}}
\put(102.00,107.67){\line(0,-1){6.67}}
\put(129.00,138.67){\line(0,-1){7.00}}
\put(129.00,127.00){\line(0,-1){14.00}}
\put(129.00,108.00){\line(0,-1){7.00}}
\put(105.33,138.33){\line(-1,1){0.33}}
\put(105.00,138.67){\line(0,0){0.00}}
\put(105.00,138.67){\line(0,-1){6.67}}
\put(126.33,138.67){\line(0,-1){7.00}}
\put(105.00,107.67){\line(0,-1){6.67}}
\put(126.33,107.67){\line(0,-1){6.67}}
\put(125.67,127.00){\line(0,-1){3.00}}
\put(125.67,124.00){\line(-5,-1){9.33}}
\put(117.33,119.67){\line(5,1){8.33}}
\put(126.33,121.00){\line(0,-1){8.33}}
\put(105.67,112.33){\line(0,1){3.00}}
\put(112.00,118.67){\line(-6,-1){6.33}}
\put(105.67,126.67){\line(0,-1){6.33}}
\put(112.33,121.33){\line(-4,-1){7.00}}
\put(105.67,117.33){\line(0,-1){1.67}}
\end{picture}
\label{Fig1}
\vspace{-8 cm}
\caption{Simplest tree diagrams for nucleus-nucleus scattering}
\end{figure}
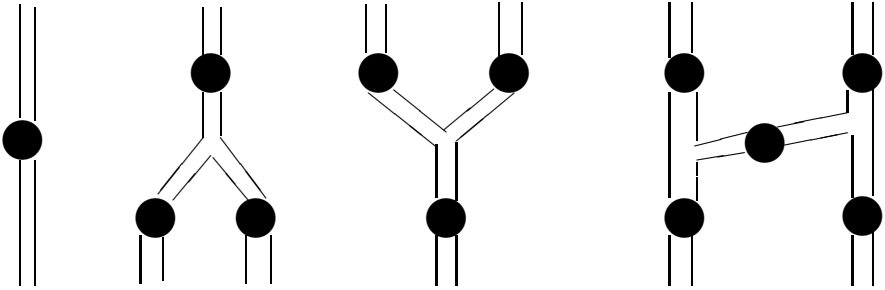
Both pomeron loops and contributions which cannot be described in terms
of pomerons (e.g. gluonic interaction between two pomerons) are
neglected, since they are damped by higher powers of $1/N_c$ and/or $1/A,1/B$.
Obviously this approximation cannot work at superlarge energies when the
pomeron can propagate to transverse distances larger than
the nuclei dimensions, which occurs at rapidities $Y$ such that
$\alpha'Y>R^2_{A,B}$, where $\alpha'$ is the pomeron slope. All experimental
data give $\alpha<0.01$ fm$^2$, so that these rapidities seem to be well
beyond our present and predictable possibilities. The approximation of
large nuclei also allows to neglect
correlations between colour distributions of different nucleons
and excludes diagrams in which a pomeron interacts simultaneously
with two nucleons of the target or projectile.

Our final result is a pair of equations for the dipole densities in the
two nuclei, which possess full conformal invariance in
absence of external sources. The equations
include terms with the interaction between the two densities.
If one neglects this interaction the equations decouple into a
pair of BK equations for the projectile and target.
The equations  are in fact very complicated and not well suited for
numerical studies,
which are difficult  already for the forward propagation case
considered in ~\cite{bra2}. Some simplification seems to be possible by
passing to the conformal basis and restricting to lowest conformal states,
which is also discussed in the paper.

\section{Pomeron diagrams and effective field theory}

At fixed overall impact parameter $b$ the AB
amplitude ${\cal A}(Y,b)$ can be presented as an
exponential of its connected part:
\beq
{\cal A}(Y,b)=2is\left(1-e^{- T(Y,b)}\right).
\eeq
The dimensionless $T$ is an
integral over
two impact parameters $b_A$ and $b_B$ of the collision point relative to
the centers of the nuclei A and B:
\beq
T(Y,b)=\int d^2b_Ad^2b_B\delta^2(b-b_A+b_B)T(Y, b_A,b_B).
\eeq

As mentioned in the Introduction, in the perturbative QCD with
$N_c\rightarrow\infty$ the amplitude
$-T(Y,b_A,b_B)$ is given by a sum of all connected tree diagrams constructed
of BFKL pomerons and the triple pomeron vertex. More concretely, in these
diagrams a line ("propagator") describing propagation of a pair of gluons from
rapidity $y$
and points $r_1$ and $r_2$ to rapidity $y'$  and points $r'_1$ and $r'_2$
corresponds to  the BFKL Green function $G_{y-y'}(r_1,r_2|r'_1,r'_2)$
~\cite{lip}:
\beq
G_{y-y'}(r_1,r_2|r'_1,r'_2)=\theta(y-y')
\sum_{\mu}e^{\omega_{\mu}(y-y')}\lambda_{\mu}
E_{\mu}(r_1,r_2)
E^*_{\mu}(r'_1,r'_2),
\eeq
where $\mu=\{n,\nu,r_0\}\equiv\{h,r_0\}$, summation in (3) includes
summation over $n$ and integrations
over  $\nu$  and transverse vector $r_0$
with the weight $(\nu^2+n^2/4)/\pi^4$; functions $E_\mu$ form
the conformal basis. In the complex notation $r=x+iy$, $r^*=x-iy$
\beq
E_{\mu}(r_1,r_2)=
\left(\frac{r_{12}}{r_{10}r_{20}}\right)^{h}
\left(\frac{r^*_{12}}{r^*_{10}r^*_{20}}\right)^{\bar{h}},
\eeq
where $r_{12}=r_1-r_2$ etc; $h=(1-n)/2+i\nu$ and $\bar{h}=1-h^*$ are
conformal weights.  Function $\omega_\mu$ is the BFKL eigenvalue
\beq
\omega_{\mu}=\omega_h=
\bar{\alpha}\Big[\psi(1)-{\rm Re}\,\psi\Big(\frac{|n|+1}{2}+i\nu\Big)\Big],
\eeq
where standardly $\bar{\alpha}=\alpha_sN_c/\pi$.
Finally
\beq
\lambda_\mu=\lambda_h=\frac{1}{[(n+1)^2+4\nu^2][(n-1)^2+4\nu^2]}.
\eeq

The interaction between pomerons is realized via the triple pomeron
vertex. It is non-local and not symmetric in the incoming and outgoing
pomerons. In the limit $N_c\to\infty$ its form for the splitting of a
pomeron into two was established
in ~\cite{BV}.
The three BFKL Green functions are connected by it as  follows
(see Fig. 2a)
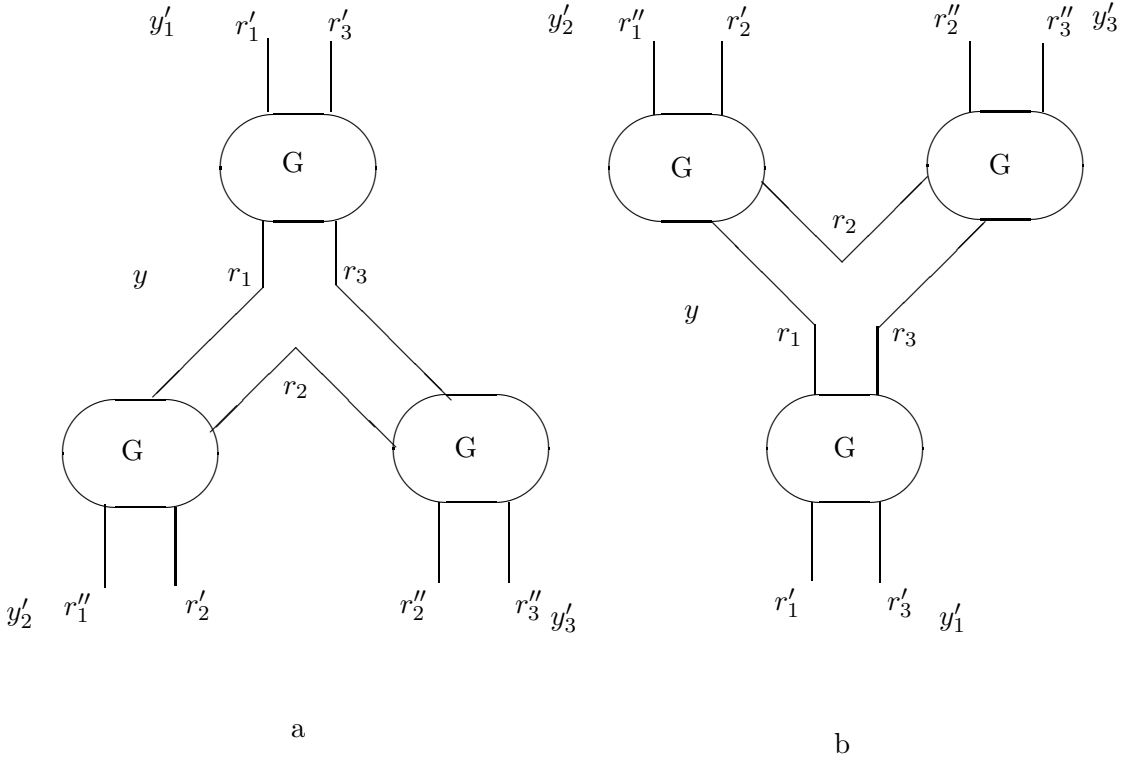
\begin{figure}
\unitlength=1.00mm
\special{em:linewidth 0.4pt}
\linethickness{0.4pt}
\begin{picture}(149.33,135.67)
\put(42.00,115.50){\oval(20.67,14.33)[]}
\put(21.00,77.50){\oval(20.67,14.33)[]}
\put(65.00,78.17){\oval(20.67,14.33)[]}
\put(93.67,115.50){\oval(20.67,14.33)[]}
\put(136.00,115.83){\oval(20.67,14.33)[]}
\put(114.67,78.17){\oval(20.67,14.33)[]}
\put(38.00,123.00){\line(0,1){9.67}}
\put(46.33,123.00){\line(0,1){9.33}}
\put(89.33,122.67){\line(0,1){9.67}}
\put(98.33,122.67){\line(0,1){9.67}}
\put(131.33,123.00){\line(0,1){9.33}}
\put(141.00,123.00){\line(0,1){9.00}}
\put(16.33,70.67){\line(0,-1){11.00}}
\put(25.67,70.33){\line(0,-1){10.33}}
\put(60.67,71.00){\line(0,-1){10.67}}
\put(70.00,71.00){\line(0,-1){10.67}}
\put(110.33,71.00){\line(0,-1){10.33}}
\put(119.33,71.00){\line(0,-1){10.67}}
\put(37.33,108.33){\line(0,-1){8.67}}
\put(47.00,108.33){\line(0,-1){8.33}}
\put(37.33,99.67){\line(-1,-1){14.67}}
\put(47.00,100.00){\line(1,-1){15.33}}
\put(41.67,91.67){\line(-1,-1){11.33}}
\put(41.67,91.67){\line(1,-1){13.33}}
\put(110.67,85.33){\line(0,1){9.33}}
\put(119.00,85.33){\line(0,1){9.00}}
\put(110.67,94.67){\line(-1,1){13.67}}
\put(119.00,94.00){\line(1,1){14.33}}
\put(114.33,103.00){\line(-1,1){10.67}}
\put(114.33,103.00){\line(1,1){11.33}}
\put(41.33,116.33){\makebox(0,0)[cc]{G}}
\put(93.00,115.67){\makebox(0,0)[cc]{G}}
\put(135.33,116.00){\makebox(0,0)[cc]{G}}
\put(20.00,100.67){\makebox(0,0)[lc]{$y$}}
\put(93.33,96.00){\makebox(0,0)[lc]{$y$}}
\put(20.00,78.00){\makebox(0,0)[cc]{G}}
\put(64.33,78.33){\makebox(0,0)[cc]{G}}
\put(114.67,78.33){\makebox(0,0)[cc]{G}}
\put(42.00,40.67){\makebox(0,0)[cc]{a}}
\put(114.33,39.00){\makebox(0,0)[cc]{b}}
\put(35.33,134.33){\makebox(0,0)[cc]{$r'_1$}}
\put(47.67,134.67){\makebox(0,0)[cc]{$r'_3$}}
\put(34.33,101.00){\makebox(0,0)[cc]{$r_1$}}
\put(49.67,101.33){\makebox(0,0)[cc]{$r_3$}}
\put(41.67,86.00){\makebox(0,0)[cc]{$r_2$}}
\put(114.67,108.00){\makebox(0,0)[cc]{$r_2$}}
\put(107.33,93.00){\makebox(0,0)[cc]{$r_1$}}
\put(122.67,93.00){\makebox(0,0)[cc]{$r_3$}}
\put(12.67,57.00){\makebox(0,0)[cc]{$r''_1$}}
\put(72.67,57.33){\makebox(0,0)[cc]{$r''_3$}}
\put(28.67,57.33){\makebox(0,0)[cc]{$r'_2$}}
\put(57.33,57.33){\makebox(0,0)[cc]{$r''_2$}}
\put(107.00,57.67){\makebox(0,0)[cc]{$r'_1$}}
\put(122.00,57.67){\makebox(0,0)[cc]{$r'_3$}}
\put(86.33,134.67){\makebox(0,0)[cc]{$r''_1$}}
\put(143.33,135.00){\makebox(0,0)[cc]{$r''_3$}}
\put(100.67,135.00){\makebox(0,0)[cc]{$r'_2$}}
\put(128.33,135.33){\makebox(0,0)[cc]{$r''_2$}}
\put(24.00,135.00){\makebox(0,0)[cc]{$y'_1$}}
\put(5.00,56.33){\makebox(0,0)[cc]{$y'_2$}}
\put(77.33,55.67){\makebox(0,0)[cc]{$y'_3$}}
\put(129.00,55.67){\makebox(0,0)[cc]{$y'_1$}}
\put(77.00,135.33){\makebox(0,0)[cc]{$y'_2$}}
\put(149.33,135.67){\makebox(0,0)[cc]{$y'_3$}}
\end{picture}
\label{Fig2}
\vspace*{-3 cm}
\caption{Splitting of a pomeron into two (a) and
fusion of two pomerons into one (b).}
\end{figure}
\beq
\frac{2\alpha_s^2N_c}{\pi}\int\frac{d^2r_1d^2r_2d^2r_3}
{r_{12}^2r_{32}^2r_{13}^2}
\Big(G_{y'_1-y} (r'_1,r'_3|r_1,r_3)L^{\dagger}_{13}\Big)\cdot
G_{y-y'_2}(r_1, r_2|r''_1,r'_2)G_{y-y'_3}( r_3,r_2|,r''_3,r''_2).
\eeq
Here $L_{13}$ is (up to a numerical factor)
the quadratic Kasimir operator for
the conformal transformations of $r_1$ and $r_3$:
\beq
L_{13}=r_{13}^4p_1^2p_3^2,
\eeq
where in the configuration space $p^2=-\nabla^2$.
In (8) $L_{13}^{\dagger}$ is acting on the left. Note that the
triple pomeron vertex is symmetric in the gluons inside the
outgoing pomerons (i.e under $r_1\leftrightarrow r_2$ and
$r_3\leftrightarrow r_2$). So the outgoing pomerons have to be
symmetric in their respective gluons.
The form of the vertex for the fusion of two pomerons into one is
actually not known. However, the symmetry between target and projectile
prompts us to assume that for the inverse process $2+3\rightarrow 1$
the BFKL functions are to be joined as    (Fig. 2b)
\beq
\frac{2\alpha_s^2N_c}{\pi}\int\frac{d^2r_1d^2r_2d^2r_3}{r_{12}^2r_{32}^2
r_{13}^2}
G_{y'_2-y}(r''_1, r'_2|r_1,r_2)G_{y'_3-y}( r''_3,r''_2|r_3,r_2)
L_{13}G_{y-y'_1}(r_1,r_3|r'_1,r'_3).
\eeq

Finally we have to describe the interaction of the pomerons with the
two nuclei. The BFKL Green functions corresponding to the external
legs of the diagrams are to be integrated with the colour density of
each nucleus. We take the target nucleus (B) at rest, that is, at rapidity
zero. Then each outgoing external BFKL Green function is to be transformed
into
\beq
\int d^2r_1'd^2r'_2 G_y(r_1,r_2|r'_1,r_2')\rho_B(r'_1,r'_2)\equiv
\int dy'd^2r'_1d^2r'_2G_{y-y'}(r_1,r_2|r'_1,r_2')\tau_B(y',r'_1,r'_2),
\eeq
where $\rho_B$ is the colour density of the target. If we neglect
correlations between the colours of the nucleons in the nucleus B
\beq
\tau_B(y,r_1,r_2)=g^2BT_B(b_B)\delta^2\left(b_B-\frac{r_1+r_2}{2}\right)
\rho_N(r_{12})\delta(y),
\eeq
where $T_B$ is the
profile function of the nucleus B and
$\rho_N$ is the colour density of the nucleon.
Similarly each ingoing BFKL external
Green function is to be transformed into
\beq
\int dy'd^2r'_1d^2r'_2\tau_A(y',r'_1,r_2)G_{y'-y}(r'_1,r'_2|r_1,r_2),
\eeq
where
\beq
\tau_A(y,r_1,r_2)=g^2AT_A(b_A)\delta^2\left(b_A-\frac{r_1+r_2}{2}\right)
\rho_N(r_{12})\delta(y-Y)
\eeq
and $Y$ is the overall rapidity.
If the densities $\tau_A$ and $\tau_B$ are symmetric in the gluons
(which is true for (11) and (13) and will be assumed in the following) then
the initial pomerons are also symmetric in the gluons inside them.
Together with the mentioned properties of the triple pomeron verteces
it means that only symmetric pomeron states are propagating in the
two nuclei.

To find the amplitude, one has to sum over all connected
diagrams with $M$ ingoing and
$N$ outgoing lines, corresponding to $M$ interactions with the projectile
and $N$ interactions with the target, divided by $M!N!$.

It is trivial to see that this sum exactly corresponds to the sum of
tree diagrams generated by an effective quantum theory of two pomeronic
fields (dipole densities, up to a factor)
$\Phi(y,r_1,r_2)$ and $\Phi^{\dagger}(y,r_1,r_2)$, symmetric in
$r_1,r_2$, with the
action
\[
S=S_0+S_I+S_E
\]
consisting of three terms, which correspond to free pomerons,
their mutual interaction and their
interaction with external sourses (nuclei) respectively.

To give the correct propagators $S_0$ has to be chosen as
\beq
S_0=\int dydy'd^2r_1d^2r_2d^2r'_1d^2r'_2\Phi^{\dagger}(y,r_1,r_2)
G^{-1}_{y-y'}(r_1,r_2|r'_1,r'_2)\Phi(y',r'_1,r'_2)\equiv
\langle\Phi^{\dagger}|G^{-1}|\Phi\rangle,
\eeq
where $\langle|\rangle$ means the integration over $y$ and both gluon
coordinates.
Note that the sign of $S_0$ corresponds to the following substitution
of the conventionally defined field variables $\Phi$ and $\Phi^{\dagger}$:
\beq
\Phi\rightarrow i\Phi,\ \ \Phi^{\dagger}\rightarrow i\Phi^{\dagger},
\eeq
which allows to make all terms of the action real.

According to (7) and (9) the interaction term $S_I$ is local in rapidity
\beq
S_I=
\frac{2\alpha_s^2N_c}{\pi}\int dy
\int\frac{d^2r_1d^2r_2d^2r_3}{r_{12}^2r_{32}^2r_{13}^2}
\Big\{ \Big(L_{13}\Phi(y,r_1,r_3)\Big)\cdot\Phi^{\dagger}(y,r_1,r_2)
\Phi^{\dagger}(y, r_3,r_2)
\ \ +\ \ h.c.\ \Big\}.
\eeq
The overall sign combines the initial factor $i$ and $i^3$ from the
substitution (15).

Finally the interaction with the nuclei is local both in rapidity and
coordinates:
\beq
S_E=-\int dyd^2r_1d^2r_2
\Big\{\Phi(y,r_1,r_2)\tau_A(y,r_1,r_2)+\Phi^{\dagger}(y,r_1,r_2)
\tau_B(y,r_1,r_2)
\Big\}.
\eeq
The minus sign comes from the initial $i$ and the substitution (15).

The amplitude $T(Y,b_A,b_B)$ is then expressed through a functional
integral
\beq
Z=\int D\Phi D\Phi^{\dagger}e^{S}.
\eeq
In the tree approximation, corresponding
to diagrams of the type shown in Fig. 1, keeping only connected diagrams
we find
\beq
T(Y,b_A,b_b)=-\ln \frac{Z}{Z_0}=-S\{\Phi,\Phi^{\dagger}\}
\eeq
where $Z_0 (=1)$ is the value of $Z$ for $S_E=0$.
and the action  $S$ is to be calculated for $\Phi$ and $\Phi^{\dagger}$
satifying the classical equation of motion.

\section{Equations for the dipole densities}
The classical equations of motion follow from the variation of the action
with respect to  $\Phi$ and $\Phi^{\dagger}$:
\beq
\frac{\delta S}{\delta \Phi(y,r_1,r_2)}=
\frac{\delta S}{\delta \Phi^{\dagger}(y,r_1,r_2)} =0.
\eeq
We find a pair of equations
\[
G^{-1}\Phi(y,r_1,r_2)+
\frac{2\alpha_s^2N_c}{\pi}\int \frac{d^2r_3}
{r_{12}^2r_{32}^2r_{13}^2}
\Big\{\Phi(y,r_1,r_3)\Phi(y,r_2,r_3)L_{12}
\]\beq+
2\Phi^{\dagger}(y,r_3,r_2)L_{13}\Phi(y,r_1,r_3)
\Big\}=
\tau_B(y,r_1,r_2)
\eeq
and
\[
\Phi^{\dagger}(y,r_1,r_2)G^{-1}+
\frac{2\alpha_s^2N_c}{\pi}\int \frac{d^2r_3}
{r_{12}^2r_{32}^2r_{13}^2}
\Big\{\Phi^{\dagger}(y,r_1,r_3)
\Phi^{\dagger}(y,r_2,r_3)L_{12}
\]\beq+
2\Phi(y,r_3,r_2)L_{13}\Phi^{\dagger}(y,r_1,r_3)
\Big\}=
\tau_A(y,r_1,r_2),
\eeq
where the operators $L_{12}$ are assumed to act on the left.
These equations have also to be supplemented by conditions
\beq
\Phi(y,r_1,r_2)=0\ \ {\rm if}\ \ y<0,\ \
\Phi^{\dagger}(y,r_1,r_2)=0\ \ {\rm if}\ \ y>Y.
\eeq
To write the equations in their final form we  note that the BFKL Green
function $G$ satisfies the equations
\beq
p_1^2p_2^2\Big(\frac{\partial}{\partial y}+H\Big)G=
\Big(\frac{\partial}{\partial y}+H^{\dagger}\Big)p_1^2p_2^2G=1,
\eeq
where $H$ is the BFKL Hamiltonian:
\beq
H=\frac{\bar{\alpha}}{2}\Big(\ln p_1^2+\ln p_2^2 +
\frac{1}{p_1^2}\ln r_{12}^2\cdot p_1^2+
\frac{1}{p_2^2}\ln r_{12}^2\cdot p_2^2-4\psi(1)\Big).
\eeq
Fom this we conclude that
\beq
G^{-1}=p_1^2p_2^2\Big(\frac{\partial}{\partial y}+H\Big)=
\Big(\frac{\partial}{\partial y}+H^{\dagger}\Big)p_1^2p_2^2.
\eeq
Multiplying Eqs. (21)and (22) by $p_1^{-2}p_2^{-2}$ fom the left
and from the right respectively we finally find
\[
\Big(\frac{\partial}{\partial y}+H\Big)\Phi(y,r_1,r_2)+
\frac{2\alpha_s^2N_c}{\pi}\int \frac{d^2r_3r_{12}^2}
{r_{32}^2r_{13}^2}
\Phi(y,r_1,r_3)\Phi(y,r_2,r_3)
\]\beq+
\frac{4\alpha_s^2N_c}{\pi}L_{12}^{-1}\int \frac{d^2r_3r_{12}^2}
{r_{32}^2r_{13}^2}
\Phi^{\dagger}(y,r_3,r_2)L_{13}\Phi(y,r_1,r_3)
=
\tau_B(y,r_1,r_2)
\eeq
and
\[
\Big(-\frac{\partial}{\partial y}+H\Big)\Phi^{\dagger}(y,r_1,r_2)+
\frac{2\alpha_s^2N_c}{\pi}\int \frac{d^2r_3r_{12}^2}
{r_{32}^2r_{13}^2}
\Phi^{\dagger}(y,r_1,r_3)\Phi^{\dagger}(y,r_2,r_3)
\]\beq+
\frac{4\alpha_s^2N_c}{\pi}L_{12}^{-1}\int \frac{d^2r_3r_{12}^2}
{r_{32}^2r_{13}^2}
\Phi(y,r_3,r_2)L_{13}\Phi^{\dagger}(y,r_1,r_3)
=
\tau_A(y,r_1,r_2).
\eeq
The $\delta$-like dependence of the external sources on $y$ together
with conditions (23) imply that one can drop the sources in the
equations and substitute them by the boundary conditions at $y=0$ and $y=Y$:
\beq
\Phi(y,r_1,r_2)_{y=0}=\rho_B(r_1,r_2),\ \
\Phi^{\dagger}(y,r_1,r_2)_{y=Y}=\rho_A(r_1,r_2),
\eeq
where $\rho_{A,B}$ are given by (11) and (13) without the
$\delta$-functions in rapidity.

From the form of equations one immediately concludes that they are
conformal invariant provided the external sources possess this invariance.

These equations can be also written in the form which allows an easy
comparison with the BK equation for non-forward fan diagrams.
To this aim one
rescales the fields putting
\beq
 \Phi(r_1,r_2)=\frac{N(r_1,r_2)}{4\pi\alpha_s},\ \
\Phi^{\dagger}(r_1,r_2)=\frac{N(r_1,r_2)}{4\pi\alpha_s}
\eeq
and uses a representation for the Hamiltonian $H$  ~\cite{BLV}
\beq
Hf(r_1,r_2)=
\frac{\bar{\alpha}}{2\pi}\int\frac{d^2r_3r_{12}^2}
{ r_{23}^2r_{13}^2}\Big(f(r_1,r_2)-f(r_1,r_3)-f(r_2,r_3)\Big).
\eeq
Then our equations take the form
\[
\frac{\partial N(r_1,r_2)}{\partial y}=
-\frac{\bar{\alpha}}{2\pi}\int\frac{d^2r_3r_{12}^2}
{ r_{23}^2r_{13}^2}\Big(N(r_1,r_2)-N(r_1,r_3)-N(r_2,r_3)
+N(r_1,r_3)N(r_2,r_3)\Big)\]\beq
-\frac{\bar{\alpha}}{\pi}L_{12}^{-1}\int \frac{d^2r_3r_{12}^2}
{r_{32}^2r_{13}^2}
N^{\dagger}(r_3,r_2)L_{13}N(r_1,r_3)
=0
\eeq
and
\[
\frac{\partial N^{\dagger}(r_1,r_2)}{\partial y}=
\frac{\bar{\alpha}}{2\pi}\int\frac{d^2r_3r_{12}^2}
{ r_{23}^2r_{13}^2}\Big(N^{\dagger}(r_1,r_2)-N^{\dagger}(r_1,r_3)-
N^{\dagger}(r_2,r_3)
+N^{\dagger}(r_1,r_3)N^{\dagger}(r_2,r_3)\Big)\]\beq
+\frac{\bar{\alpha}}{\pi}L_{12}^{-1}\int \frac{d^2r_3r_{12}^2}
{r_{32}^2r_{13}^2}
N(r_3,r_2)L_{13}N^{\dagger}(r_1,r_3),
=0
\eeq
with the boundary conditions which follow from (29) after  rescaling (30).

If one neglects the second term on the r.h.s in both equations and thus
decouples $N$ and $N^{\dagger}$ the equations turn into a pair of independent
BK equations for dipole densities in the nuclei A and B evolving in
opposite directions in rapidity and corresponding to the sum of two sets
of fan diagrams starting from the projectile or target. However the second
terms introduce interaction between these sets and correspond to diagrams
which contain both splitting and fusion of pomerons. The structure of this
interaction is rather complicated in both configuration and momentum spaces
due to non-locality of the inverse operator $L^{-1}$. One expects it to be
simplified in the conformal basis, which will be the subject of the next
section.

Meanwhile, using the equations of motion one can simplify the expression
for the action $S$ calculated on their solution.
Indeed multiplying Eqs. (21) and (22) by $\Phi^{\dagger}(y,r_1,r_2)$ and
$\Phi(y,r_1,r_2)$, integrating over $y, r_1,r_2$ and summing the results
one obtains   a relation
\beq
2S_0+3S_{I}+S_{E}=0.
\eeq
This can be used to exclude one of the parts of the action when
calculating the amplitude $T$. Recalling that the fields are discontinuous
at the boundaries we obtain from (34)
\beq
T(Y,b_A,b_B)=\frac{1}{3}(S_E-S_0)=\frac{1}{2}(S_I-S_E).
\eeq

\section{Equations in the conformal basis}
One may hope that the equations for the gluon densities
may be somewhat simplified in the conformal basis
formed by functions $E_\mu(r_1,r_2)$. To this end we present
\beq
\Phi(y,r_1,r_2)=\sum_\mu E_{\mu}(r_1,r_2)\Phi_{\mu}(y).
\eeq
The orthonormalization properties of $E_{\mu}(r_1,r_2)$
~\cite{lip} allow to invert this relation and find
\beq
\Phi_{\mu}(y)=\int\frac{d^2r_1d^2r_2}{r_{12}^4}E_{\mu}^*(r_1,r_2)
\Phi(y,r_1,r_2).
\eeq
Since $\mu=\{n,\nu,r_0\}$, transition to the conformal basis by itself
does not change the number of variables (three). However it drastically
simplifies the operators $L$ in the the mixing term of our equations.

Indeed the mixing term of Eq. (27) can be written as
\beq
T^{mix}(r_1,r_2)=\frac{4\alpha_s^2N_c}{\pi}
L_{12}^{-1}\int \frac{d^2r_3r_{12}^2}
{r_{32}^2r_{13}^2}
\sum_{\mu_1,\mu_2}\Phi_{\mu_1}^{\dagger}(y)\Phi_{\mu_2}(y)
\lambda_{\mu_2}^{-1}
E_{\mu_1}^*(r_3,r_2)E_{\mu_2}(r_1,r_3),
\eeq
where we have used that
\beq
L_{13}E_{\mu}(r_1,r_3)=\lambda_{\mu}^{-1}E_{\mu}(r_1,r_3).
\eeq
Expanding the integral over $r_3$ considered as a function of
$r_1$ and $r_2$
in the conformal basis we get
\beq
T^{mix}(r_1,r_2)=\sum_{\mu}T^{mix}_{\mu}E_{\mu}(r_1,r_2),
\eeq
where according to (39)
\[
T_{\mu}^{mix}=\frac{4\alpha_s^2N_c}{\pi}
\int\frac{d^2r_1d^2r_2}{r_{12}^4}E_{\mu}^*(r_1,r_2)
\]\beq
\times L_{12}^{-1}\int \frac{d^2r_3r_{12}^2}
{r_{32}^2r_{13}^2}
\sum_{\mu_1,\mu_2}\Phi_{\mu_1}^{\dagger}(y)
\Phi_{\mu_2}(y)\lambda_{\mu_2}^{-1}
E_{\mu_1}^*(r_3,r_2)E_{\mu_2}(r_1,r_3).
\eeq
We integrate by parts transforming action of $L_{12}^{-1}$ on
$E_{\mu}^*(r_1,r_2)/r_{12}^4$ and use
\beq
r_{12}^4{L_{12}^{-1}}^{\dagger}r_{12}^{-4}= L_{12}^{-1}
\eeq
to apply $L_{12}^{-1}$ directly on $E_{\mu}(r_1,r_2)$ which gives a factor
$\lambda_{\mu}$.
So in the end we get
\[
T_{\mu}^{mix}=\frac{4\alpha_s^2N_c}{\pi}
\int\frac{d^2r_1d^2r^2d^2r_3}
{r_{12}^2r_{32}^2r_{13}^2}\lambda_{\mu}E_{\mu}^*(r_1,r_2)
\sum_{\mu_1,\mu_2}
\Phi_{\mu_1}^{\dagger}(y)\Phi_{\mu_2}(y)
\lambda_{\mu_2}^{-1}
E_{\mu_1}^*(r_3,r_2)E_{\mu_2}(r_1,r_3)\]\beq=
\frac{4\alpha_s^2N_c}{\pi}\lambda_{\mu}\sum_{\mu_1,\mu_2}
V_{\tilde{\mu},\tilde{\mu}_1,\mu_2}\lambda_{\mu_2}^{-1}
\Phi_{\mu_1}^{\dagger}(y)\Phi_{\mu_2}(y),
\eeq
where $V_{\mu,\mu_1,\mu_2}$ is the triple pomeron vertex in the
conformal basis
\beq
V_{\mu,\mu_1,\mu_2}=
\int\frac{d^2r_1d^2r^2d^2r_3}
{r_{12}^2r_{32}^2r_{13}^2}E_{\mu}(r_1,r_2)
E_{\mu_1}(r_3,r_2)E_{\mu_2}(r_1,r_3)
\eeq
and $\tilde{\mu}$ corresponds to the complex conjugate basis function:
if $\mu=\{h,r_0\}$ then $\tilde{\mu}=\{1-h,r_0\}$
(and always $\bar{h}=1-h^*$).

A similar transformation of the first integral term in Eq. (27) is
straightforward and leads to the result which is different from (43)
by the absence of conjugate $\Phi$'s and factors $\lambda$.
So we find the first equation in the conformal basis as
(suppressing the common argument $y$ and dropping the source term)
\beq
\frac{\partial\Phi_{\mu}}{\partial y}=\omega_{\mu}\Phi_{\mu}
-\frac{2\alpha_s^2N_c}{\pi}\sum_{\mu_1,\mu_2}\Phi_{\mu_2}
\Big(V_{\tilde{\mu},\mu_1,\mu_2}\Phi_{\mu_1}+2
\frac{\lambda_{\mu}}{\lambda_{\mu_2}}V_{\tilde{\mu},\tilde{\mu}_1,\mu_2}
\Phi_{\mu_1}^{\dagger}\Big)
\eeq
The second equation can be obtained by reversing the direction of propagation
in rapidity and passing to conjugate fields:
\beq
\frac{\partial\Phi_{\mu}^{\dagger}}{\partial y}=
-\omega_{\mu}\Phi_{\mu}^{\dagger}
+\frac{2\alpha_s^2N_c}{\pi}\sum_{\mu_1,\mu_2}\Phi_{\mu_2}^{\dagger}
\Big(V_{\mu,\tilde{\mu}_1,\tilde{\mu}_2}\Phi_{\mu_1}^{\dagger}+
2\frac{\lambda_{\mu}}{\lambda_{\mu_2}}
V_{\mu,\mu_1,\tilde{\mu}_2}\Phi_{\mu_1}\Big).
\eeq

The triple pomeron vertex  $V_{\mu,\mu_1,\mu_2}$ was studied in
~\cite{kor}. It depends on three conformal weights $h,h_1$ and $h_2$ and
three center-of-mass vectors $\{r_0, r_{01},r_{02}\}
\equiv\{\rho_0,\rho_1,\rho_2\} $. The dependence on the latters is determined
by the conformal invariance, so that  (in complex notation)
\beq
V_{\mu,\mu_1,\mu_2}=\Omega_{h,h_1,h_2}\prod_{i<j}\rho_{ij}^{-\Delta_{ij}}
{\rho^*_{ij}}^{-\bar{\Delta}_{ij}},
\eeq
where $i=0,1,2$, $\Delta_{01}=h_0+h_1-h_2$, $\bar{\Delta}_{01}=
\bar{h}_0+\bar{h}_1-\bar{h}_2$ etc. The part of the vertex depending
on conformal weights $\Omega_{h,h_1,h_2}$ was found in ~\cite{kor}
for arbitrary conformal weights in terms of the Meijer function
$G^{pq}_{44}$.
The complicated form of the vertex together with the use of complex
variables make equations (45), (46) for the gluon densities in the general
case not very suitable for practical calculations, in spite of the
simplification for the action of operators $L$. However they may
serve as a starting point for further simplifications realized by
truncating the equations by certain low values of conformal weights.
In the next section we consider a most drastic example of such a
truncation.

\section{Lowest conformal weights}
As well-known from the study of the linear BFKL equation in the
high-energy limit, the leading contribution comes from the minimal
conformal weight in the expansion (36), namely $h=\bar{h}=1/2$, which
corresponds to $n=0$ and $\nu=0$. So the simplest case which may be
of interest for our problem is to put $n=0$ in all places and $\nu=0$
whenever this is allowed by the equations, that is in $\Phi_\mu$,
$\Phi_\mu^{\dagger}$ and $\Omega_{h,h_1h_2}$. Then one finds for the
triple pomeron coupling ~\cite{kor}
\beq
\Omega_{1/2,1/2,1/2}\equiv\Omega_0=2\pi^7{_4F_3}(1/2){_6F_5}(1/2)=7766.679.
\eeq
The unknown fields $\Phi_{\mu}$ and $\Phi^{\dagger}_{\mu}$ become functions
of rapidity $y$ and center-of mass vector $\rho_0$ (actually of $\rho_0^2$
due to rotational invariance).
The two equations (45) and (46) simplify to
\[
\frac{\partial\Phi(\rho_0)}{\partial y}=\]\beq\Delta\Phi(\rho_0)
-\frac{\alpha_s^2N_c}{8\pi^7}
\Omega_0\int\frac{d^2\rho_1d^2\rho_2}{\rho_{01}\rho_{02}}
\delta''\Big(\ln\frac{\rho_{01}}{\rho_{02}\rho_{12}}\Big)
\delta''\Big(\ln\frac{\rho_{02}}{\rho_{01}\rho_{12}}\Big)
\Phi(\rho_1)\Big(\Phi(\rho_2)+2\Phi^{\dagger}(\rho_2)\Big)
\eeq
and
\[
\frac{\partial\Phi^{\dagger}(\rho_0)}{\partial y}=\]\beq-\Delta\Phi^{\dagger}(\rho_0)
+\frac{\alpha_s^2N_c}{8\pi^7}
\Omega_0\int\frac{d^2\rho_1d^2\rho_2}{\rho_{01}\rho_{02}}
\delta''\Big(\ln\frac{\rho_{01}}{\rho_{02}\rho_{12}}\Big)
\delta''\Big(\ln\frac{\rho_{02}}{\rho_{01}\rho_{12}}\Big)
\Phi^{\dagger}(\rho_1)\Big(\Phi^{\dagger}(\rho_2)+2\Phi(\rho_2)\Big),
\eeq
where $\Delta=\omega_{n=0,\nu=0}$ is the BFKL intercept.
We have taken into account that due to the presence of the $\delta$-
functions we have $\rho_{12}=1$ (in the chosen scale, determined by the
sources).

From the assumed independence of the fields of $\nu$ it follows that the
boundary conditions for these equations have to belong to the class of
functions of the form
\beq
f(r_1,r_2)=\int\frac{r_{12}d^2r_3}{r_{12}r_{13}}
\delta''\Big(\ln\frac{r_{12}}{r_{13}r_{23}}\Big)g(r_3),
\eeq
where $g(r)$ is an arbitrary function.  Obviously this restricts the sources
to be of a very special sort, with the dependence on two vectors
$r_1$ and $r_2$ and thus on three variables $r_1^2, r_2^2$ and
$\bf{r}_1\bf{r}_2$ determined by a function of a single variable $r_3^2$.
So any practical use of the thus simplified system of equations is
questionable. At most it may serve to study the qualitative features of
the solution in the limit of high energies, when one may hope that the
influence of the choice of the boundary conditions becomes neglegible
(as it happens with the BK equation). Still even Eqs. (49) and (50) do not
look easily solvable. We reserve their study for a separate publication.

\section{Conclusions}
We have derived a pair of equations for dipole densities in two heavy nuclei,
which describe their
scattering in the perturbative QCD with a large number of colours.
The equations contain mixing terms which are
both non-linear and non-local. In absence of mixing the equations
decouple into a pair of BK equations for the projectile and target.

In contrast to the hA case the
equations are
to be solved with given boundary conditions at rapidities of
the projectile and target, which complicates their solution enormously.
The equations themselves are conformal invariant. This invariance is
naturally broken by the sources. However use of the conformal basis
may open ways for various simplifications of the equations, which may
facilitate their solution, if only on the qualitative level.

\section{Acknowledgements}
This work was supported by the NATO Grant PST.CLG.980287

\end{document}